\documentclass[aps,preprint]{revtex4}%
\usepackage{amsfonts}
\usepackage{amsmath}
\usepackage{amssymb}
\usepackage{graphicx}%
\begin{document}
\preprint{ }
\title{Speed of Light in Gravitational Fields}
\author{Yukio Tomozawa}
\email{tomozawa@umich.edu}
\affiliation{The Michigan Center for Theoretical Physics and Randall Lab. of Physics,
University of Michigan, Ann Arbor, MI. 48109-1120}
\date{Revised January 23, 2004}

\begin{abstract}
A spherically symmetric and static metric that describes physical coordinates
is introduced. It is defined to be a metric that gives coordinate independent
results for physically observable quantities without a further coordinate
transformation. The suggested metric also makes a prediction for the second
order gravitational red shift effect that can be utilized for a precision
experimental test in the future. A possible new experimental test would be
provided by a modern Michelson Morley experiment on the earth with the two
arms in vertical and horizontal directions to see the validity of isotropy or
anisotropy for the speed of light. The possibility of using pulsars and GPS
(Global Positioning System) for a general relativity test is discussed.

\end{abstract}
\pacs{04.20.-q, 04.80.-y, 04.80.Cc, 95.30.Sf}
\maketitle

\section{Introduction}

The coordinates in the Schwartzschild metric or the Eddington isotropic metric
do not correspond to physically observable coordinates in the gravitational
field of a mass. One should specify the relationship between these coordinates
and physically observable coordinates in order to compare observations with
the theoretical predictions of general relativity. In this article, we
introduce a physical metric as one in which the coordinates in the metric
represent observable coordinates, by showing that a coordinate independent
result is reproduced by the physical metric. In particular, the time delay
experiment is found to be crucial for the determination of the physical
metric, while all other experimental tests of general relativity that have
been done in the past are insensitive to the choice of the metric in the first
order of gravity correction.

\section{The physical metric}

The physical metric for a spherically symmetric and static point mass $M$,%

\begin{equation}
ds^{2}=e^{\nu(r)}dt^{2}-e^{\lambda(r)}dr^{2}-e^{\mu(r)}r^{2}(d\theta^{2}%
+\sin^{2}\theta d\phi^{2}), \label{eq0}%
\end{equation}
is related to the Schwartzschild metric,%

\begin{equation}
ds^{2}=(1-r_{s}/r^{\prime})dt^{2}-(1/(1-r_{s}/r^{\prime}))dr^{\prime
2}-\ r^{\prime2}(d\theta^{2}+\sin^{2}\theta d\phi^{2}), \label{eqa}%
\end{equation}
by the transformation, $r^{\prime}=re^{\mu(r)/2},$ where $r_{s}=2GM/c^{2}$ is
the Schwartzschild radius. Then one gets%

\begin{equation}
e^{\nu(r)}=1-(r_{s}/r)e^{-\mu(r)/2}=1+a_{1}(r_{s}/r)+a_{2}(r_{s}/r)^{2}%
+\cdots,
\end{equation}

\begin{equation}
e^{\lambda(r)}=(\frac{d}{dr}(re^{\mu(r)/2}))^{2}/(1-(r_{s}/r)e^{-\mu
(r)/2})=1+b_{1}(r_{s}/r)+b_{2}(r_{s}/r)^{2}+\cdots,
\end{equation}

\begin{equation}
e^{\mu(r)}=1+c_{1}(r_{s}/r)+c_{2}(r_{s}/r)^{2}+\cdots,
\end{equation}
where asymptotic expansion is used and%

\begin{equation}
a_{1}=-1,\text{ }and\ \ a_{2}=c_{1}/2, \label{eq7}%
\end{equation}

\begin{equation}
b_{1}=1,\text{ }and\text{ \ }b_{2}=1-c_{1}/2+c_{1}^{2}/4-c_{2}.
\end{equation}
To first order of $\ r_{s}/r$ , the metric is expressed as%

\begin{equation}
e^{\nu(r)}=1-r_{s}/r+\cdots,\text{ }e^{\lambda(r)}=1+r_{s}/r+\cdots,\text{
}and\text{\ \ }e^{\mu(r)}=1+c_{1}(r_{s}/r)+\cdots.
\end{equation}
All the experimental tests of general relativity so far can be expressed in
terms of this metric.

\section{The geodesic equations}

The geodesic equations can be obtained from variations of the line integral
over an invariant parameter $\tau$,$\ \int(\frac{ds}{d\tau})^{2}d\tau$, and
their integrals are given by%

\begin{equation}
\frac{dt}{d\tau}=e^{-\nu(r)}, \label{eq1}%
\end{equation}

\begin{equation}
\frac{d\phi}{d\tau}=J_{\phi}e^{-\mu(r)}/(r\sin\theta)^{2},
\end{equation}

\begin{equation}
(\frac{d\theta}{d\tau})^{2}=(J_{\theta}^{\text{ }2}-J_{\phi}^{\text{ }2}%
/\sin^{2}\theta)e^{-2\mu(r)}/r^{4}.
\end{equation}
Restricting the plane of motion to $\frac{d\theta}{d\tau}=0,$ $\theta=\pi/2,$
the radial part of the geodesic integral is given by%

\begin{equation}
(\frac{dr}{d\tau})^{2}=e^{-\lambda(r)}(e^{-\nu(r)}-J^{\text{ }2}e^{-\mu
(r)}/r^{2}-E) \label{eq2}%
\end{equation}
where $J_{\phi}$, $J_{\theta}$ and $E$ are constants of integration and%

\begin{equation}
J^{\text{ }2}=J_{\phi}^{\text{ }2}=J_{\theta}^{\text{ }2}.
\end{equation}
for the above restriction on the plane of motion. The constant $E$ is 0 for
light propagation.

\section{Time delay and speed of light}

From Eq. (\ref{eq1}) and Eq. (\ref{eq2}), it follows that%
\begin{align}
\frac{dt}{dr}  &  =\pm\text{ }e^{-\nu(r)}/\sqrt{e^{-\nu(r)-\lambda
(r)}-J^{\text{ }2}e^{-\mu(r)-\lambda(r)}/r^{2}}\\
&  =\pm\text{ }\frac{r}{\sqrt{r^{2}-r_{0}^{\text{ }2}}}\text{ }(1+\frac
{(b_{1}-a_{1})\text{ }r_{s}}{2r}+\frac{(c_{1}-a_{1})\text{ }r_{0}\text{ }%
r_{s}}{2\text{ }r\text{ }(r+r_{0})}+\cdots)
\end{align}
for light propagation, where $r_{0}$ is the impact parameter. Integrating from
the distance between the planet and the sun, $r_{1}$, to the distance between
the earth and the sun, $r_{2\text{ }}$, one gets the expression for the time
delay experiment of Shapiro et.al. \cite{shapiro 1} (for the return trip of
the light),%
\begin{equation}
\bigtriangleup t=2\text{ }r_{s}\text{ }(\ln(\frac{r_{1}+\sqrt{r_{1}^{\text{
}2}-r_{0}^{\text{ \ }2}}}{r_{2}-\sqrt{r_{2}^{\text{ \ }2}-r_{0}^{\text{ \ }2}%
}})+\frac{c_{1}+1}{2}(\sqrt{\frac{r_{1}-r_{0}}{r_{1}+r_{0}}}+\sqrt{\frac
{r_{2}-r_{0}}{r_{2}+r_{0}}})). \label{eq3}%
\end{equation}
(See the ref. \cite{wein} for the calculation for the Schwartzschild metric,
$c_{1}=0$.)

The second term of Eq. (\ref{eq3}) depends on the choice of the value of
$c_{1}$ and can be eliminated by a further coordinate transformation,%
\begin{equation}
r=r^{\prime\prime}e^{\mu(r^{\prime\prime})/2}=1+c_{1}^{\prime\prime}%
/2(r_{s}/r^{\prime\prime})+\cdots.
\end{equation}
Therefore, the coordinate independent prediction of general relativity should
be%
\begin{equation}
\bigtriangleup t=2\text{ }r_{s}\text{ }\ln(\frac{r_{1}+\sqrt{r_{1}^{\text{ }%
2}-r_{0}^{\text{ \ }2}}}{r_{2}-\sqrt{r_{2}^{\text{ \ }2}-r_{0}^{\text{ \ }2}}%
})\label{eqshapiro1}%
\end{equation}
This is the result also obtained by the PPN (Post Newtonian Method)\cite{mtw},
and agrees with the most recent observational data \cite{shapiro 2} with high
accurracy (1 in 1000 accurracy). By comparing Eqs. (\ref{eq3}) and
(\ref{eqshapiro1}), we conclude that the physical metric is determined by the
condition,
\begin{equation}
c_{1}=-1.\label{eqc1=-1}%
\end{equation}
We note that the parameter values%
\begin{equation}
a_{1}=-1,\text{ }and\text{ }b_{1}=1
\end{equation}
are coordinate indepependent and determined by being the solution of the
Einstein equation and the physical bounbary condition. Thus we conclude that
Eq. (\ref{eqc1=-1}) is the condition for the physical metric.

On the other hand, the coordinate speed of light in the gravitational field
represented by the physical metric is obtained as%
\begin{equation}
c_{g}=\sqrt{(\frac{dr}{dt})^{2}+(r\frac{d\phi}{dt})^{2}}=e^{\nu(r)}%
\sqrt{(e^{-\nu(r)-\lambda(r)}-J^{2}e^{-\mu(r)-\lambda(r)}/r^{2})+J^{2}%
e^{-2\mu(r)}/r^{2}},
\end{equation}
where\ $J^{2}=r_{0}^{\text{ \ }2}e^{\mu(r_{0})-\nu(r_{0})}$. Using asymptotic
expansion, one gets%
\begin{align}
c_{g} &  =1-\frac{r_{s}}{r}\text{ }((\frac{b_{1}-a_{1}}{2})(1-(\frac{r_{0}}%
{r})^{2})+(\frac{c_{1}-b_{1}}{2})\text{ }(\frac{r_{0}}{r})^{2})+\cdots\\
&  =1-\frac{r_{s}}{r}(\cos^{2}\omega+(\frac{c_{1}+1}{2})\sin^{2}\omega
)+\cdots,\label{eqc3}%
\end{align}
where $\omega$ is the angle between the direction of the source of gravity and
the direction of the propagation of light. Here the bending of light can be
neglected as a higher order correction, and $(\frac{r_{0}}{r})^{2}=\sin
^{2}\omega$. \ 

With the choice, Eq. (\ref{eqc1=-1}), one gets
\begin{equation}
c_{g}=1-\frac{r_{s}}{r}\cos^{2}\omega+\cdots, \label{eqcg1}%
\end{equation}
for the coordinate speed of light. This is consistent with the coordinate
speed of light obtained by the condition, $ds^{2}=0.$ If one uses the local
time,%
\begin{equation}
d\tau_{p}=e^{\nu(r)/2}dt, \label{proper}%
\end{equation}
the coordinate speed of light becomes%

\begin{align}
c_{g}^{\prime} &  =c_{g}\frac{dt}{d\tau_{p}}=c_{g}(1+\frac{r_{s}}{2r}%
+\cdots)\\
&  =1-\frac{r_{s}}{2r}\text{ }(2\cos^{2}\omega-1)+\cdots=1-\frac{r_{s}}%
{2r}\cos2\omega+\cdots\label{eqcg2}%
\end{align}

The question remains what is the observable speed of light in an environment
of gravity such as on the earth. If one defines the speed of light by%
\begin{equation}
c_{g}^{\prime\prime}=\sqrt{e^{\lambda(r)}(\frac{dr}{d\tau_{p}})^{2}+e^{\mu
(r)}(r\frac{d\phi}{d\tau_{p}})^{2}}=e^{\nu(r)/2}\sqrt{(e^{-\nu(r)}%
-J^{2}e^{-\mu(r)}/r^{2})+J^{2}e^{-2\mu(r)}/r^{2}}=1\label{isotropy}%
\end{equation}
This implies that in coordinates for which the radial length, $dr$, is
streched as $(1+$ $\frac{r_{s}}{2r}+\cdots)dr$ and the angular length,
$rd\phi$, is shortened as $(1-\frac{r_{s}}{2r}+\cdots)rd\phi$ , the speed of
light is equal to 1 ($=c$). It is the author's opinion that this statement
does not correspond to the observable speed of light. This can be seen in the
following manner. Suppose one tries to make a modern Michelson Morley
experiment \cite{laser}by using two laser cavities, one in a horizontal
direction and the other in the vertical direction. Prepare two identical
cavities lying in the horizontal direction. By bringing one of the cavities to
the vertical direction, its length is shortened by the force of gravity. If
the material of the cavities has a very high Young's modulus, the lengths of
the two cavities are almost identical. Then, a modern Michelson Morley
experiment with this instrument (with an appropriate correction for the
gravitational shrinkage effect) should show a difference in the speed of
light  based on Eq. (\ref{eqcg1}) or Eq. (\ref{eqcg2}), but not based on Eq.
(\ref{isotropy}).

\section{The other experimental tests}

The other tests of general relativity are shown to be insensitive to the
presence of the $c_{1}$ term. For the bending of light, one uses the formula,%
\begin{align}
\frac{d\phi}{dr} &  =\pm\text{ }e^{-\mu(r)+\lambda(r)/2}/r^{2}\sqrt
{e^{-\nu(r)}/J^{2}-e^{-\mu(r)}/r^{2}}\\
&  =\pm\frac{r_{0}}{r\sqrt{r^{2}-r_{0}^{2}}}(1+r_{s}(\frac{b_{1}}{2r}%
-\frac{a_{1}r}{2r_{0}(r+r_{0})}+\frac{c_{1}}{2}(\frac{r}{r_{0}(r+r_{0})}%
-\frac{1}{r}))+\cdots).\label{eq6}%
\end{align}
Integrating this from a large distance, one gets the well known expression for
the bending of light,%
\begin{equation}
\triangle\phi=(b_{1}-a_{1})\text{ }\frac{r_{s}}{r_{0}}=\frac{2\text{ }r_{s}%
}{r_{0}}.\label{eq8}%
\end{equation}
The integration of the $c_{1}$ term in Eq. (\ref{eq6}) gives a vanishingly
small value and therefore insensitive to the value of $c_{1}$, as is seen from
Eq. (\ref{eq8}).\cite{duff}

For the advancement of perihelia, one uses the formula%
\begin{align}
\frac{d\phi}{dr} &  =\pm\text{ }e^{-\mu(r)+\lambda(r)/2}/r^{2}\sqrt
{e^{-\nu(r)}/J^{2}-e^{-\mu(r)}/r^{2}-E}\\
&  =\pm\frac{1}{r^{2}\sqrt{(\frac{1}{r_{\cdot}}-\frac{1}{r})(\frac{1}{r}%
-\frac{1}{r_{+}})}}(1+\frac{r_{s}}{2}(\frac{b_{1}}{r}+(-a_{1}+\frac{a_{2}%
}{a_{1}})\frac{r_{+}+r_{\cdot}}{r_{+}r_{\cdot}}+c_{1}(\frac{1}{r_{+}}+\frac
{1}{r_{\cdot}}-\frac{1}{r}))+\cdots)
\end{align}
where $r_{\pm}$ are the semi major and minor axis of the elliptical orbit. The
appearance of \ $a_{2}$ is necessitated by the cancellation of the lowest term
for the determinatin of the constants $J^{2}$ and \ $E/J^{2}$. Integration
over the ellipse yields the advancement of perihelion,%
\begin{equation}
\triangle\phi=\frac{\pi r_{s}}{2}(\frac{1}{r_{+}}+\frac{1}{r_{-}}%
)(b_{1}-2a_{1}+c_{1}+\frac{2a_{2}}{a_{1}}).
\end{equation}
Due to the relationship, Eq. (\ref{eq7}), $c_{1}+\frac{2a_{2}}{a_{1}}=0$, one
obtains%
\begin{equation}
\triangle\phi=\frac{\pi r_{s}}{2}(\frac{1}{r_{+}}+\frac{1}{r_{-}}%
)(b_{1}-2a_{1})=\frac{3\pi r_{s}}{2}(\frac{1}{r_{+}}+\frac{1}{r_{-}%
}).\label{eq9}%
\end{equation}
It is remarkable that the $c_{1}$ term and $a_{2}$ term cancel each other and
the final result is again independent of $c_{1}$.\cite{duff}. In other words,
both equations, Eq. (\ref{eq8}) and Eq. (\ref{eq9}), which have been supported
by observational data, are insensitive to the value of $c_{1}$. The reason for
these phenomena is that the bending of light and the advancement of perihelia
are variations in the angular variables, which are less ambiguous coordinates.
On the other hand, the time delay experiment, Eq. (\ref{eq3}), and the speed
of light, Eq. (\ref{eqc3}), formally depend on the parameter $c_{1}.$

\section{Suggested experimental tests}

In the following, the author suggests possible experiments of various types.

(i) Pulsar time delay experiments: In order to improve the statistics of time
delay experiments, the author suggests doing a time delay experiment on
pulsars with small declination angles in ecliptic coordinates. Such pulsars
cross, graze the sun or nearly do so once a year and provide an opportunity to
perform the experiment. Candidates for such pulsars are listed with J-names
and the J2000 ecliptic RA and DE in parenthses (in degrees) in the order of
small DE angle\cite{pulsarcat}: J1022+10 (153.864, -0.06982), J0540+2329
(86.139, 0.10214), J1744-2334 (266.495, -0.17392), J1817-2312 (273.915,
0.18425), J1730-2304 (263.186, 0.19150), J1800-2343 (270.012, -0.27921),
J1801-2451 (270.221, 0.33398), J1801-2316 (270.306, 0.33426), J1822-2256
(275.290, 0.38682), J1733-2228 (263.866, 0.82122), J0614+2229 (93.299,
-0.89891), J1757-2421 (269.4470, -0.92775), J0629+2415 (96.629, 0.98950\}.
Since the angular size of the radius of the sun is 0.267 degrees, the first 5
pulsars in the list cross the sun. The time delay experiments by binary
pulsars have been performed.\cite{taylor}

(ii) Speed of light experiments:

A recent series of laser beat experiments, which is called modern
Michelson-Morley experiments, tests the isotropy of the speed of light in
horizontal directions with good accuracy (on the order of $\delta
c/c\simeq10^{-15}$).\cite{laser} It is desirable to do a modern
Michelson-Morley experiment with the two arms in vertical and horizontal
directions in order to see the effect of the earth's gravity on the variation
of the speed of light or absence of it, as was suggested earlier in this
article. Since the characteristic parameter at the surface of the earth is
$r_{s}/r=1.39\ast10^{-9}$ and a simple minded application of the
Schwartzschild metric gives anisotropy for speed of light of this order, Eq.
(\ref{eqc3}) with $c_{1}=0$, it is worthwhile to examine isotropy or
anisotropy in this type with the accuracy of $10^{-10}.$

(iii) Use of GPS. It is known that distance (or time) measurement by GPS has
errors of the order of a few meters due to the atmospheric index of refraction
and other factors. It is therefore essential to reduce the errors in order to
perform a general relativity experiment with GPS. One possible suggestion is
to use the measurement of a LEO (Low Earth Orbit) satellite to subtract the
effect of the free electron density and get a distance measurement between a
GPS and a LEO satellite. There is still the remaining effect of the upper
atmosphere to be eliminated. An alternative test of general relativity by GPS
would be a time dilation test like the Pound-Rebka experiment\cite{poundr}.
The time difference between the atomic clocks on a GPS and on the ground (such
as NIST) is recorded as a monitoring operation. Subtracting the effects of the
Doppler shifts from the GPS motion and the earth's rotation, one can get the
time delay by (general and special) relativity. This is equivalent to the
Pound Rebka experiment. The advantage of the GPS experiment is that it can
improve statistics by a continuous operation.

(iv) A second order test of the gravitational red shift. Using the physical
metric derived in this article, one can derive a formula for gravitational red
shift in second order in the gravitational constant. From Eq. (\ref{eq7}), Eq.
(\ref{eqc1=-1}) and Eq. (\ref{proper}), it follows that%
\begin{align}
d\tau_{p} &  =\sqrt{1+a_{1}(r_{s}/r)+a_{2}(r_{s}/r)^{2}+\cdots}dt\\
&  =\sqrt{1-(r_{s}/r)-(r_{s}/r)^{2}/2+\cdots}dt.
\end{align}
Here, the use of the physical metric enables us to get a coordinate
independent prediction based on%
\begin{equation}
a_{2}=c_{1}/2=-1/2.
\end{equation}
This prediction can be utilized as a second order test of general relativity
when a precise measurement of the gravitational red shift becomes available in
the future. One possible direction is the measurement of spectral red shifts
from binary white dwarfs.

\section{Discussions}

Some discussions are due. Eq. (\ref{eq0}) can be generalized to%

\begin{equation}
ds^{2}=e^{\nu(r)}dt^{2}+2e^{\kappa(r)}dtdr-e^{\lambda(r)}dr^{2}-e^{\mu
(r)}r^{2}(d\theta^{2}+\sin^{2}\theta d\phi^{2}).
\end{equation}
This introduces an extra parameter in the theory,%
\begin{equation}
e^{\kappa(r)}=d_{1}(r_{s}/r)+\ \cdots.
\end{equation}
It is easy to see that the only change in first order in Eq. (\ref{eqa})
through Eq. (\ref{eq2}) is in the geodesic integral%
\begin{equation}
e^{\nu(r)}\frac{dt}{d\tau}=1-e^{\kappa(r)}\frac{dr}{d\tau}.\label{eq20}%
\end{equation}
The change in the geodesic integral for the radial coordinate is in a second
order of $r_{s}/r$ , while those for the angular coordinates are invariant. As
a result, one obtains the change for Eq. (\ref{eqcg1}),%
\begin{equation}
c_{g}=1-\frac{r_{s}}{r}(\cos^{2}\omega+d_{1}\cos\omega)+\cdots.
\end{equation}
An additional term for time delay experiments cancels for the return trip or
gives an unobservable constant shift for a oneway trip in pulsar time delay
experiments. The predictions for the bending of light and the advancement of
perihelia are not affected by the change in Eq. (\ref{eq20}). In other words,
the final result for the rest of discussion in this article is unchanged.

Finally, the author emphsizes that Eq. (\ref{eq3}) or Eq. (\ref{eqc3})
provides a challenge for new experimental tests of general relativity.

\begin{acknowledgments}
It is a great pleasure to thank David N. Williams for reading the manuscript
and Mike J. Duff, Peter G. Halverson, Jean Krish, James T. Liu, John D.
Monnier, J. Keith Riles, Bin Wang and David N. Williams for useful discussion.
\end{acknowledgments}

\end{document}